# Deciphering transcriptional dynamics in vivo by counting nascent RNA molecules

Sandeep Choubey (1), Jane Kondev (1), Alvaro Sanchez (2)

(1) Department of Physics, Brandeis University, Waltham, Massachusetts 02453,
(2) Rowland Institute at Harvard, Harvard University, Cambridge, Massachusetts 02142

Transcription of genes is the focus of most forms of regulation of gene expression. Even though careful biochemical experimentation has revealed the molecular mechanisms of transcription initiation for a number of different promoters in vitro, the dynamics of this process in cells is still poorly understood. One approach has been to measure the transcriptional output (fluorescently labeled messenger RNAs or proteins) from single cells in a genetically identical population, which could then be compared to predictions from models that incorporate different molecular mechanisms of transcription initiation. However, this approach suffers from the problem, that processes downstream from transcription can affect the measured output and therefore mask the signature of stochastic transcription initiation on the cell-to-cell variability of the transcriptional outputs. Here we show theoretically that measurements of the cell-to-cell variability in the number of nascent RNAs provide a more direct test of the mechanism of transcription initiation. We derive exact expressions for the first two moments of the distribution of nascent RNA molecules and apply our theory to published data for a collection of constitutively expressed yeast genes. We find that the measured nascent RNA distributions are inconsistent with transcription initiation proceeding via one rate-limiting step, which has been generally inferred from measurements of cytoplasmic messenger RNA. Instead, we propose a two-step mechanism of initiation, which is consistent with the available data. These findings for the yeast promoters highlight the utility of our theory for deciphering transcriptional dynamics in vivo from experiments that count nascent RNA molecules in single cells.

## Introduction

Transcription is a multi-step process that leads to the production of messenger RNA (mRNA) molecules from its DNA template. Genetic experiments on cells have identified the key molecular components of transcription, while biochemical studies with purified components have uncovered the basic mechanisms governing their interactions. Still an important question that remains is whether the same mechanisms are also operational in cells.

One approach to unraveling the mechanisms of transcription in cells is to measure the outputs of this process, either the proteins that correspond to the genes being transcribed, or the actual mRNA molecules. This idea has motivated numerous experiments that count protein (1–3), and mRNA (4–6) molecules in single cell and the measured steady state distribution of these molecules in a clonal cell population have been used to infer the dynamics of transcription (4, 5). For instance, analysis of the steady state distributions of cytoplasmic mRNA in yeast for a number of different genes, revealed two distinct mechanisms of transcription initiation: one-step and ON-OFF (4). One-step initiation is characterized by stochastic production and degradation of mRNA that is exponentially distributed in time. ON-OFF initiation assumes that the promoter stochastically switches between an ON and an OFF state. Transcription proceeds stochastically only from the ON state and the resulting mRNA is eventually degraded.

While this approach to deciphering transcriptional dynamics in cells has been very fruitful, a key limitation is that, processes that are downstream from transcription can mask the signature of stochastic transcriptional dynamics in measurements of the cell-to-cell variability of mRNA and protein abundances. A striking example of this is the recent finding that spatial and temporal averaging, i.e., the process of accumulation and diffusion of mRNA transcripts during nuclear cycles, significantly reduces the variability in mRNA copy number expected from stochastic transcription initiation (7). In addition, effects such as mRNA transport out of the nucleus, mRNA processing, and nonlinear mRNA degradation (8–12) can also in principle alter the level of variability of cytoplasmic mRNA. All of these non-transcriptional sources of variability may propagate to the protein level as well, affecting the cell-to-cell fluctuations in protein copy number, in addition to the stochastic nature of translation. Finally, it has been recently shown that partitioning of both mRNA and protein molecules during cell division (13–15) can generate noise patterns similar to those that would be generated by stochastic transcription and translation. Therefore, the cell-to-cell variability of both protein and cytoplasmic mRNA copy number do not necessarily reflect transcriptional dynamics alone but are determined by a combination of stochastic processes of which transcriptional dynamics is just one component (16).

One alternative to analyzing steady state mRNA and protein distributions, has been to directly image transcription in real time using fluorescently labeled RNA binding proteins that associate with nascent RNA, which is still in the process of being assembled at the gene by the RNA polymerase (17–21). When applied to E. coli, Dyctostelium or animal cells this technique revealed widespread transcriptional bursting consistent with the ON-OFF mechanism of transcription initiation. In contrast, in experiments on constitutive genes in *S.cerevisiae*, Larson *et al.* (21) found that the transcription initiation process is dominated by one rate limiting step. In spite of the great promises of this approach, it is technically challenging and still remains in its infancy.

Lately, a score of experimental papers have reported measurements of distributions of nascent RNA transcripts at a single gene and across a clonal cell population, using single-molecule FISH (4, 22–24). These experiments reveal the number of RNA polymerases engaged in transcribing a single gene in a single cell at a specific instant in time. This information can also be obtained from so-called Miller spreads (electron micrographs of intact chromosomes extracted from cells) which reveal transcribing polymerases along different genes that can be counted (25–29). Counting nascent RNAs (or the number of transcribing polymerases) provides a more direct readout for the transcriptional dynamics at the promoter within the short window of time required for an RNA polymerase molecule to complete elongation (for a typical gene in yeast the elongation time is of the order of few minutes (4)). As such, this experimental approach is not affected by the aforementioned stochastic processes that contribute to cytoplasmic mRNA and protein fluctuations. Indeed, as mentioned above, strong discrepancies between cytoplasmic and nascent mRNA distributions have been recently found in Drosophila embryos (7). Below we also demonstrate similar discrepancies in conclusions reached from counting nascent and cytoplasmic mRNA in yeast cells.

Experiments that count the number of nascent RNAs in individual cells call for mathematical models that can connect molecular mechanisms of transcription initiation with measured nascent RNA distributions. One of the key results that we report here is the development of such a model. In particular, we show how to compute the mean and variance of the distribution of nascent RNAs for an arbitrary mechanism of transcription initiation. The results of these calculations provide the tools to extract information about transcriptional dynamics from nascent RNA distributions. We demonstrate the usefulness of our method by analyzing published nascent RNA distributions for a set of constitutively expressed yeast genes (23). By applying our model to this data we find that these yeast genes have similar average initiation rates. We also find that initiation of transcription of these yeast genes is a two-step process, where the average durations of both steps are equal. This is in sharp contrast to the conclusion that was reached from counting cytoplasmic mRNAs for some of these genes, namely that transcription initiation is dominated by one rate limiting step (23).

**Results**

**Distribution of the number of nascent RNAs is determined by the dynamics of transcription initiation and elongation**

In order to connect mechanisms of transcription initiation with nascent RNA distributions, we consider a model of transcriptional dynamics with an arbitrarily complex initiation mechanism followed by an elongation process. We describe both processes using chemical master equations. This approach is inspired by the work of Kepler et al. (30) who computed the moments of the mRNA distribution for a ON-OFF promoter, where the promoter slowly switches between an active (ON) and an inactive (OFF) state. Sanchez et al. (31–33) developed this method further to compute the moments of mRNA and protein distributions for arbitrarily complex promoters that can switch between multiple states, each state leading to transcript production at a particular rate. Here we implement the same master equation approach to compute the first and second moments of the nascent RNA distribution.

The promoter switches between different states as different transcription factors bind and fall off their respective binding sites, causing the effective initiation rate to fluctuate. We assume that after initiation, each RNA polymerase (RNAP) moves along the gene by hopping from one to the next base at a constant probability per unit time (Fig. 1A). Our model assumes that transcription initiation timescales are much slower than the elongation timescale and hence RNAPs do not interfere with each other while moving along the gene. This approximation is reasonable for all but the strongest promoters characterized by very fast initiation (34, 35). We demonstrate this explicitly using numerical simulations (36, 37) which include a detailed model of transcription elongation that takes into account excluded-volume interaction between adjacent polymerases, as well as RNAP pausing (Fig. S1A). The agreement between analytical results based on our simple model and the simulations of the more realistic model that incorporates traffic and pausing of RNAPs only starts to break down when the initiation time scales become comparable to the elongation time scales (Figs. S1C-D). We conclude that for typical rates reported for RNAP elongation and pausing the simple model of transcription adopted here reproduces the first two moments of the nascent RNA distribution with deviations from those obtained from the more realistic model that are less than 10% as long as initiation of transcription is slower than 30 initiations/min. All the initiation rates that have been reported so far from in vivo measurements are slower(4, 17, 21), with important exceptions such as the ribosomal promoters (34, 35). Hence, despite its simplicity, the model of transcription initiation and elongation we adopt here should apply to most genes in *E.Coli* and yeast.

In order to compute the first two moments of the nascent RNA distribution for an arbitrary transcription initiation mechanism, we consider a promoter that can exist in $N$ possible states. The rate of transition from the $s$-th to the $q$-th state is $k_{s,q}$, and the rate at which RNAP initiates transcription from the $s$-th promoter state is $k_{s,ini}$. Following the initiation process, every RNAP elongates by hopping from one base to the next with a probability per unit time (which is equal to the average rate of elongation) $k$. The number of RNAP molecules, which is the same as the number of nascent RNAs, at the $i$-th base pair is $m_i$. Hence the number of nascent RNAs ($M$) along a gene of length $L$ bases, is given by, $M = \sum_{i=1}^{L} m_i$. The state of the promoter+gene system is described by ($L$+1) stochastic variables: the number of nascent RNAs ($m_1,...,m_L$) at every base along the gene, and the label $s$, characterizing the state of the promoter. Hence the probability distribution function that characterizes the promoter+gene system is given by $P(s,m_1,...,m_L)$. To stream-line the mathematics we define the following probability vector:

$$\vec{P}(m_1,...,m_L) = (P(1,m_1,...,m_L), P(2,m_1,...,m_L),...,P(s,m_1,...,m_L)) \ . \tag{1}$$

The time evolution for this probability vector can be described by chemical master equations, which can be written in compact, matrix form as

$$\frac{d\vec{P}(m_1,...,m_L)}{dt} = (\hat{K} - \hat{R} - \hat{\Gamma}\sum_{i=1}^{L} m_i) \vec{P}(m_1,.,m_i,.,m_L) + \hat{R} \vec{P}(m_1-1,...,m_L)$$

$$+ \sum_{i=1}^{L} k(m_i+1) \hat{\Gamma} \vec{P}(m_1,.,m_i+1,m_{i+1}-1,.,m_L) \ . \tag{2}$$

In Eq. 2, we define the following matrices: $\hat{K}$, which describes the transition between different promoter states, and whose elements are $K_{ss} = k_{q,s}$ if $q \neq s$ and $K_{ss} = -\sum_{q} k_{q,s}$. $\hat{R}$ is a matrix that contains the rates of initiation from different promoter states. In the case of one-step initiation it is diagonal with the diagonal elements being the rates of initiation from different promoter states. In the case of two-step initiation this matrix is off-diagonal owing to the fact that the promoter state changes after initiation (for details please see in the *SI*[1]). $\hat{\Gamma}$ is also diagonal and its elements represent the hopping rate for the polymerase from one base pair to the next, i.e., $\Gamma_{sq} = k\delta_{s,q}$.

We limit our calculation to the steady state of the nascent RNA distribution for which the left hand side of Eq. 2 is set to zero. To obtain the first and second moments of the number of nascent RNAs, $M = \sum_{i=1}^{L} m_i$ in steady state we use Eq. 2 to compute the quantities, $\langle m_i \rangle$ and $\langle m_i m_j \rangle$ for all $i, j \leq L$. Even though the random variables $m_i$ for different bases $i$ on the gene are mutually dependent, we end up deriving a set of linear equations for $\langle m_i \rangle$ and $\langle m_i m_j \rangle$ (see *SI*). We find that these equations for the moments close, in other words they do not depend on any further, higher moments of the $m_i$'s. These linear equations can then be solved to obtain exact expressions for the first two moments of *M* as a function of all the rates that define the molecular mechanism of initiation under investigation. (For the mathematical details please see the *SI*).

**Different mechanisms of transcription initiation can be discriminated by the nascent RNA distributions they produce**

In order to demonstrate how the distribution of nascent RNAs at the transcription site can be used to extract dynamical information about the process of transcription initiation in vivo, we consider a canonical model of transcription shown in Fig. 1A (30). The promoter can switch between two states: an active state (ON), from which transcription initiation can occur, and an inactive state (OFF) from which initiation does not occur. The two states might correspond to a free promoter and one bound by a repressor protein, or a promoter uncovered or covered by a nucleosome. In theoretical studies to date transcription initiation from the ON state was typically assumed to be dominated by one rate-limiting step. Instead of initiation being a one-step process we consider the possibility that there are two rate-limiting steps involved in transcription initiation from the active state. These could be the assembly of the transcriptional machinery at the promoter (in prokaryotes, this would correspond to the formation of open complex by RNAP (38–40)), which occurs with a rate $k_{OPEN}$, followed by the RNA polymerase escaping the promoter into an elongation state (with rate $k_{ESC}$). Three different limits of this model correspond to models of transcription initiation that have been previously explored in the literature (4, 17, 41–43).

First we consider the limit when the promoter is always active ($k_{OFF} \to 0$ in Fig. 1A) and initiation is a one-step process. This is a situation when one of the two kinetic steps leading up to initiation (either the assembly of the transcriptional machinery or the escape of RNA

---

[1] Supplementary Information (SI) is attached to the main text, after the Reference section

polymerase from the promoter) is much slower than the other. In this case we find the variance equal to the mean of the nascent RNA distribution.

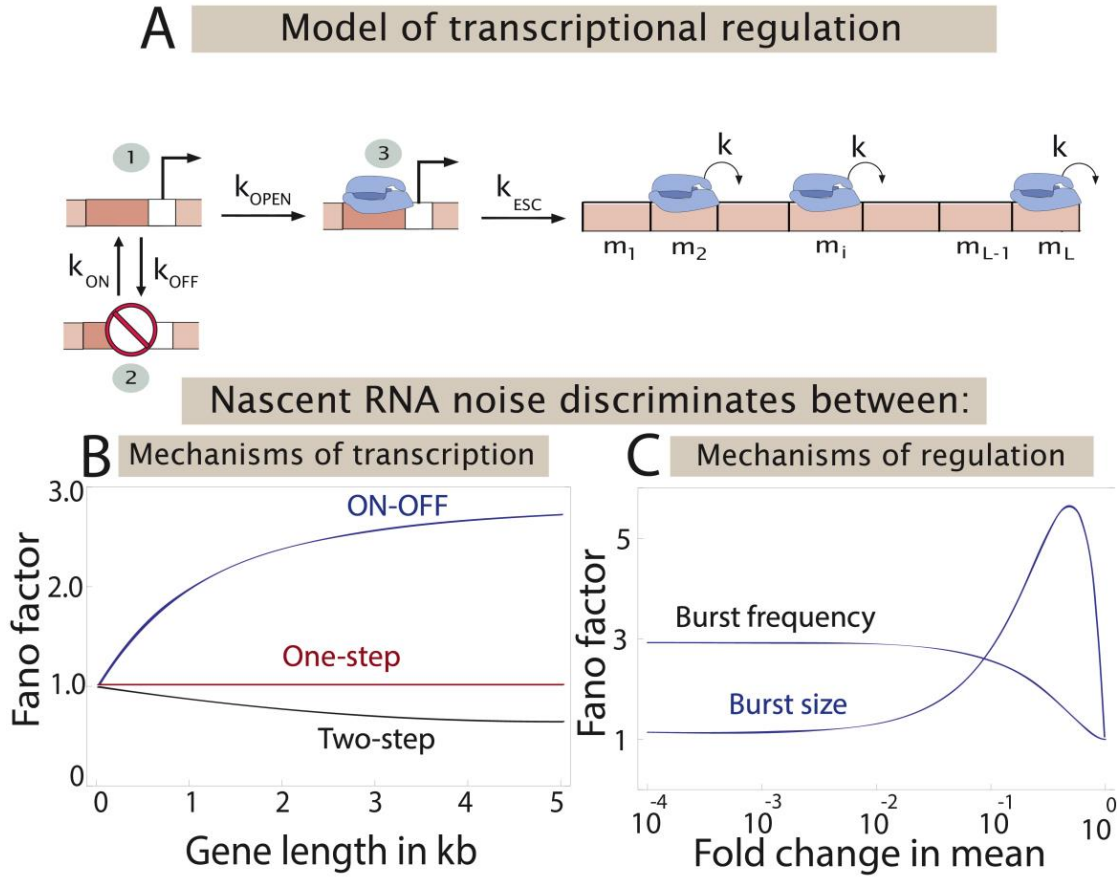

Fig. 1. (A) **Model of transcriptional regulation:** The promoter switches between two states: an active (ON) and an inactive one (OFF). The probability per unit time of switching from the active state to the inactive state is $k_{OFF}$, and from the inactive to the active state it is $k_{ON}$. Transcription initiation does not occur from the inactive state. From the active state transcription initiation occurs in two sequential steps: the formation of the pre-initiation complex at the promoter (in bacteria, this would represent formation of open complex by RNA polymerase) proceeds with rate $k_{OPEN}$ after which the RNA polymerase escapes the promoter at a constant probability per unit time $k_{ESC}$. Once on the gene the polymerases move from one base pair to the next with a rate $k$, until they reach the end of the gene and they fall off with the same rate. From this model we compute the mean and the variance of the number of RNA polymerases that are present on the gene in steady state, as a function of all the rates and the length of the gene $L$. This calculation is aided by introducing the $m_i$ variables for every base, which keep track of the number of polymerases at that base. (B) **Noise profile for different models of transcription initiation.** From the master equation of the model described in (A) we computed the Fano factor (ratio of the variance and the mean) of the nascent RNA distribution as a function of the length of the gene being transcribed, for the three different models of transcription initiation: one-step (red), ON-OFF (blue) and two-step initiation (black). The three different models give qualitatively distinct predictions. To illustrate this point for the ON-OFF model we use the following parameters: $k_{OFF}$=5/min, $k_{ON}$=0.435/min , $k$=0.8kb/min, $k_{OPEN}$= 5/min and $k_{ESC}$=0/min, which are characteristic of the PDR5 promoter in yeast, as reported in (4). For the two-step model we use $_{OPEN}$= 0.14/min, $k_{ESC}$=0.14/min,

$k_{OFF}$=0/min, $k_{ON}$=0/min , $k$=0.8kb/min, characteristic of MDN1 promoter, which we find by analyzing the data reported in ref.(23). For the one-step model, we use $k_{OPEN}$= 0.09/min, $k_{ESC}$=0/min, $k_{OFF}$=0/min, $k_{ON}$=0/min , $k$=0.8kb/min, which are characteristics of the yeast gene, RPB1, obtained by analyzing the data, published in ref. (23). (C) **Noise profiles for different regulatory mechanisms.** In the ON-OFF model of transcription, the transcriptional output can be modulated either by changing the burst size or the burst frequency, which in the model can be achieved by tuning $k_{OFF}$ and $k_{ON}$. The Fano factor for the nascent RNA distribution obtained from one or the other mode of regulation is plotted as a function of the fold change in mean (i.e., the mean of the distribution normalized by the maximum number of nascent RNAs in the cell). Clearly the two modes of regulation give qualitatively distinct predictions for the noise profile. (To illustrate this point we use the following parameters: $k_{OFF}$=5/min, $k_{ON}$=0.435/min , $k$=0.8kb/min, $L$=4436 bps, $k_{INI}$= 5/min, which were reported for the PDR5 promoter in yeast (4).)

The second limit of interest is when the rates of assembly of the transcriptional machinery ($k_{OPEN}$) and promoter escape ($k_{ESC}$) have comparable magnitudes, i.e., transcription initiation is a two-step process. In this limit, transcription initiation events are anti-correlated due to the presence of a "dead-time" or refractory period in between subsequent initiation events. The third limit of interest is the ON-OFF model, when the promoter is not always active, but is slowly switching between the active and inactive states. Here we take that the initiation from the ON state proceeds in one rate-limiting step. (Note that the more complicated situation when transcription from the ON state proceeds in two steps can also be solved using the moment approach described above.)

A key prediction of our model of transcription initiation and elongation, which is described in Fig. 1A, is how the cell-to-cell fluctuations of the nascent RNA number depend on the length of the gene being transcribed. This is an interesting quantity to consider from the point of view of experiments on cells, both due to the natural variation in gene length that exists for different genes, and the ability to synthetically alter the length of the gene being expressed from a promoter of interest by genetic manipulation. Calculations of the Fano factor as a function of gene length (Fig. 1B) reveal that this quantity easily discriminates between the three models of transcription initiation described above. When the gene length is small the Fano factor is close to one for all three models of initiation. As we increase the gene length, the Fano factor increases for the ON-OFF initiation model, in which the promoter switches slowly between an active and an inactive state. The Fano factor decreases with gene length for two-step initiation, when the promoter is always active and there are two rate limiting steps leading up to elongation. Finally, in the case when initiation is dominated by one rate-limiting step and the promoter is always active, the Fano factor is equal to one, independent of gene length.

For ON-OFF promoters that switch between an active and an inactive state (for example the PDR5 gene in yeast (4)) the nascent RNA distribution can also be used to discriminate between different mechanisms of regulation. Recent experiments (44, 45) have suggested that transcriptional regulation occurs either by modulating the burst size (given by $k_{INI}/k_{OFF}$, where $k_{INI}$ = $k_{OPEN} k_{ESC}/(k_{OPEN} + k_{ESC})$ is the average rate of initiation), or by modulating the burst frequency ($k_{ON}$); it is also possible that both are tuned. In Fig. 1C we show the results of our calculations of the Fano factor for the nascent RNA distribution, using parameters that are characteristic of the PDR5 gene and assuming that transcription regulation is achieved either by tuning the burst size or the burst frequency. We see that even though both mechanisms of regulation produce Fano factors larger than one, they make qualitatively different predictions for the functional dependence of the Fano factor on the mean of the nascent RNA distribution.

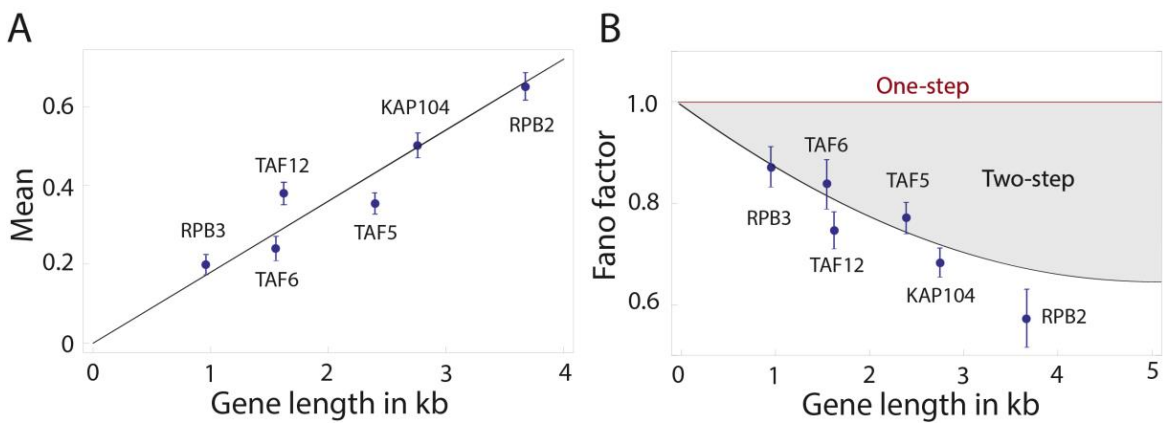

Fig. 2.(A) **Mean number of nascent RNAs for six different yeast genes.** We use the nascent RNA distribution data for six constitutively expressed yeast genes: KAP104, TAF5, TAF6, TAF12, RPB2, RPB3 and plot the mean of the distributions as a function of the gene length. The mean of the distribution increases linearly with the gene length indicating that the transcription of all six genes is initiating at the same rate. The initiation rate of these genes extracted from the data is 0.145±0.025/min, where the rate of elongation is taken to be 0.8 kb/min (4). (B) **Fano factor for the nascent RNA distribution of six different yeast genes.** Using the data for the nascent RNA distributions for the same six yeast genes described in (A) we compute the Fano factor and compare it to predictions from our mode. The shaded region shows the possible values that the Fano factor can take depending on the ratio of $k_{OPEN}$ and $k_{ESC}$ given the initiation rate determined from the mean in part (A). The boundary of the shaded region corresponds to the minimum amount of noise (as measured by the Fano factor) given the extracted rate of initiation in part (A), and it is obtained when the two rates are the same, i.e., $k_{OPEN} = k_{ESC} =$ 0.29±0.013/min. Interestingly enough the Fano factors characterizing the nascent RNA distribution for these six yeast genes seem to lie on this boundary. (The nascent RNA data for the six yeast genes used in our analysis is taken from ref. (23).)

**Measured nascent RNA distributions in yeast are consistent with a two-step mechanism of transcription initiation**

The theoretical results described above can be used as a computational tool to extract information about transcription initiation dynamics from nascent RNA distributions, which have been measured in recent experiments (4, 23). To demonstrate the utility of this approach, we analyze a set of nascent RNA distributions for twelve different constitutively expressed genes in yeast (23). We find that for six of these twelve genes (RPB2, RPB3, TAF5, TAF6, TAF12, KAP104), the mean number of nascent RNAs scales linearly with the gene length, as shown in Fig.2A. If we assume that all of these genes have comparable elongation rates (0.8 kb/min (4)), then the linear relationship between the mean nascent RNA number and gene length implies that the average initiation rates of these genes are all roughly the same and equal to 0.145±0.025/min.

In addition to the mean, our model allows us to investigate the behavior of the variance of the nascent mRNA distribution with gene length, and compare it to the predictions from different models of transcription initiation. Given that the Fano factors of the nascent RNA distribution for the six genes, RPB2, RPB3, TAF5, TAF6, TAF12, and KAP104, is less than one, the simplest model consistent with the data is one where the promoter is always active and transcription initiation is a two-step process (see Fig. 1A). This model is parameterized by the rates $k_{OPEN}$ and $k_{ESC}$. The Fano factor of the nascent RNA distribution depends on the ratio of these two rates and in Fig. 2B we show the region in the Fano factor – Gene length phase space that is consistent with the observed average initiation rate of 0.145±0.025/min for these six genes. This region (light blue shaded region in Fig. 2B) is bounded on its upper side by the limit when one of the rates is much larger than the other one (which turns initiation into a one-step process with a Fano factor equal to one), and on the lower side, by the Fano factor obtained when both rates are exactly identical. This line marks the minimum Fano factor attainable when the average initiation rate is 0.145±0.025/min. Remarkably, we find that the six genes in question have the lowest possible Fano factor. In principle, the six genes shown in Fig. 2A could lie anywhere within the shaded region in Fig. 2B. The fact that they all follow the lower boundary line indicates that these genes, which have varying length, have the same average initiation rate and have roughly the same values of $k_{OPEN}$ and $k_{ESC}$ ($k_{OPEN}$=$k_{ESC}$=0.29±0.013/min). The molecular basis of this observation remains unclear.

The remaining six (RPB1, MDN1, PUP1, PRE7, PRE3, PRP8) constitutive genes of the twelve studied (23) initiate at rates that are different than the rate of initiation that we found for the six genes discussed above (see Fig. S2). All but one of these six genes have Fano factors that are less than one, consistent with two or more steps leading up to initiation. This second set of genes thus acts as a control group that, as expected for a set of genes having different gene-specific rates of transcription, occupies the allowed region in the Fano factor-Gene length phase space without following a pattern like the one observed for the six genes discussed above (Fig. S3).

## Discussion

Direct imaging of transcriptional dynamics in real time (17–21) at the molecular scale and in individual cells still remains challenging. As an alternative, a number of recent studies have tried to decipher the dynamics of transcription initiation using the measured temporal and cell-to-cell variability of transcriptional outputs (cytoplasmic messenger RNA or protein molecules) at the single cell level. These measurements of noise have been interpreted in the context of a classification scheme for promoters, which are characterized by either a Poisson or a Gamma distribution of their outputs. These differences have then been taken to indicate a difference in the mechanism of transcription. A Poisson distribution is taken as evidence that the promoter transcribes at a constant rate, i.e., initiation is a one-step process. The Gamma distribution on the other hand is indicative of ON-OFF promoter dynamics (4). In practice, the distribution of cytoplasmic mRNA or proteins obtained from a population of cells is fitted to a mathematical model that incorporates the stochastic kinetics of transcription (and translation in the case of proteins), and the fitting parameters are interpreted as representative of the kinetic properties of stochastic gene expression (e.g., burst size, burst frequency, average transcription rate, etc.) (5). Even though in some cases this approach has produced kinetic parameters whose values are consistent with direct measurements of the same parameters (1), the interpretation of the kinetic parameters can be difficult given that the distributions of mRNA and protein may be affected by stochastic processes that occur downstream of transcription initiation. Examples of these processes include the non-linear degradation of mRNA and proteins (12), maturation time of fluorescent reporters (46), transport of mRNA out of the nucleus (8, 9), mRNA splicing (10, 11) and small RNA regulation (12, 47). Furthermore, recent theoretical results (13, 14) indicate that fluctuations due to random partitioning of molecules during cell division may yield the same mathematical dependence between noise and mean of protein and mRNA copy number in clonal cell populations, as would a stochastic model of transcription initiation and linear degradation.

An alternative to the abovementioned approaches is to count the number of transcribing polymerases (25–28), or nascent RNAs (4, 23) on the gene being transcribed, using electron micrographs and fluorescence in situ hybridization, respectively. These measurements are not affected by post-transcriptional processes and are more direct readouts of transcriptional dynamics. To date, these distributions of nascent RNAs have been used only in a qualitative or semi-quantitative way, due to the lack of mathematical models that connect these distributions with the underlying mechanisms of transcription. For instance, distributions of nascent RNAs (or of transcribing RNA polymerases) have been recently reported in yeast (4, 23, 29, 48), fly embryos (7, 48, 49), and bacteria (27, 50). The measured variances of these distributions were analyzed semi-quantitatively using simulations of particular transcription mechanisms, or simply by comparing them across samples. The model of transcription initiation and elongation developed here offers a way to quantitatively analyze measured nascent RNA distributions, and connect it to molecular mechanisms of transcription. In particular, when we consider three different models of transcription initiation that incorporate three broad classes of initiation mechanisms, we find that they make qualitatively different predictions for nascent RNA distributions.

While the one-step and ON-OFF transcription initiation mechanisms have received considerable attention in the context of transcriptional regulation (1, 7), two-step initiation mechanisms have received much less attention because of lack of experimental evidence, until very recently (41, 42). Analyzing the nascent RNA distributions for twelve constitutively expressed genes in yeast (23), we find that all but one of these distributions have Fano factors less than one. This observation is consistent with a simple model in which initiation proceeds in two-steps (for some of the genes more than two steps; see Fig. S3), which are of similar duration. The most surprising finding when analyzing these twelve genes was that six of them have not only the same average initiation rate, but also the same rates of assembly of the transcriptional machinery, and of promoter escape. More experiments, ideally ones where the dynamics of transcription are followed directly (21), would be necessary to confirm this prediction. Notably, this signature of two-step initiation is washed out at the cytoplasmic mRNA level. For instance the cytoplasmic mRNA distribution of KAP104 gene is Poisson (23), contrary to what the nascent RNA distribution suggests. It remains unclear what processes are responsible for the differences in these two distributions. In a very recent study of transcription in fly embryos, it was also found that the variability of nascent and cytoplasmic mRNA could differ by over six fold (7). In this case, the reason for this difference is spatial and temporal averaging of mRNA by diffusion and accumulation of mRNA transcripts during nuclear cycles. These two examples, demonstrate that the relationship between nascent and cytoplasmic RNA distributions is complex and context dependent.

Our findings for the yeast promoters highlight the utility of our theory for deciphering transcriptional dynamics in vivo from nascent RNA distributions. In addition, counting nascent RNAs, mRNAs and proteins simultaneously will undoubtedly further enhance our understanding of how the central dogma of molecular biology plays out in cells.

## Materials and Methods

**Model of transcriptional regulation**

To compute the first two moments of the nascent RNA distribution for the canonical model of transcriptional regulation shown in Fig. 1A we apply the general method of deriving moment equations from the master equation, Eq. 2. (For computational details please see *SI*). The rate matrices that define the master equation, Eq. 2, are in this case:

$$\hat{K} = \begin{bmatrix} -k_{ON} & k_{OFF} & 0 \\ k_{ON} & -(k_{OFF} + k_{OPEN}) & k_{ESC} \\ 0 & k_{OPEN} & -k_{ESC} \end{bmatrix},$$

$$\hat{R} = \begin{bmatrix} 0 & 0 & 0 \\ 0 & 0 & k_{ESC} \\ 0 & 0 & 0 \end{bmatrix},$$

$$\hat{\Gamma} = \begin{bmatrix} k & 0 & 0 \\ 0 & k & 0 \\ 0 & 0 & k \end{bmatrix}.$$

Where $\hat{K}$ is the transition matrix, which describes promoter switching between the three possible states shown in Fig. 1A. When an RNA polymerase initiates transcription from the promoter state that has the polymerase bound to the promoter, the state of the promoter changes to the state in which the promoter does not have a bound polymerase. This accounts for the rate of escape appearing in the transition matrix and also for $\hat{R}$ (the initiation rate matrix) not being diagonal. Using these matrices in the master equation for the nascent RNA distribution (Eq. 1) we compute analytically the mean and the variance of the distribution as a function of the gene length $L$. These results were then used to make the plots in Figs. 1B and 1C.

**Parameter selection**

We generated the plots for the Fano factor versus gene length (Fig. 1B), for the three limits of the model in Fig. 1A using the parameters listed below. For the ON-OFF promoter, where the promoter slowly switches between inactive and inactive states, we use $k_{OFF}$=5/min, $k_{ON}$=0.435/min , $k$=0.8kb/min, $k_{OPEN}$= 5/min and $k_{ESC}$=0/min; $k_{OFF}$, $k_{ON}$, $k$, and $k_{OPEN}$ are the characteristic rates for the PDR5 promoter, as reported in (4). For the two-step initiation model, where the promoter does not switch between an active and an inactive state but has two rate limiting steps leading up to initiation, we use $k_{OPEN}$= 0.14/min, $k_{ESC}$=0.14/min, $k_{OFF}$=0/min, $k_{ON}$=0/min , $k$=0.8kb/min; these are characteristics of yeast genes, such as MDN1 (23). For the one-step model, there is one rate limiting step leading up to transcription elongation and we choose $k_{OPEN}$= 0.09/min, $k_{ESC}$=0/min, $k_{OFF}$=0/min, $k_{ON}$=0/min , $k$=0.8kb/min, which are characteristics of the yeast gene, RPB1 (23).

Genes that are expressed from a promoter that switches between an ON and an OFF state can be regulated by changing the rates of switching between these two states, either by modulating the burst size (given by $k_{INI}/k_{OFF}$, where $k_{INI}$ = $k_{OPEN} k_{ESC}/(k_{OPEN} + k_{ESC}$ is the average rate of initiation), or by modulating the burst frequency ($k_{ON}$), (it is also possible that both are modulated) (44, 45). In order to examine the predictions for the nascent RNA distribution for these two mechanisms of regulation in Fig. 1C, we change burst size and burst frequency by changing $k_{OFF}$ and $k_{ON}$. In the first case, we change the burst size by changing $k_{OFF}$ and taking the other parameter values to be, $k_{ON}$=0.435/min , $k$=0.8kb/min, $L$=4436 bps, $k_{INI}$= 5/min as reported for PDR5 (4). Then we change burst frequency by changing $k_{ON}$, where the other parameters are, $k_{OFF}$=5/min, $k$=0.8kb/min, $L$=4436 bps, $k_{INI}$= 5/min as reported for PDR5 (4). In Fig. 1C the Fano factor of the nascent RNA distribution is plotted as function of its mean normalized by mean$_{max}$, where mean$_{max}$ is the maximum of the mean number of nascent RNAs which is obtained when there is no regulation and the promoter is always active.

## ACKNOWLEDGMENTS

We wish to thank Rob Phillips, Hernan Garcia, Jeff Gelles, Timothy Harden for years of stimulating discussions and shared thoughts about transcriptional dynamics. We also want to express our gratitude to Saumil Gandhi for sharing data and for illuminating discussions. A.S. was funded by a Junior Fellowship from the Rowland Institute at Harvard. This work was also supported by the NSF through grant DMR-1206146 to J.K.

# SUPPLEMENTARY INFORMATION

# Deciphering transcription dynamics in vivo by counting nascent RNA molecules

Sandeep Choubey, Jane Kondev, Alvaro Sanchez

## The first two moments of the nascent RNA distribution

In order to connect mechanisms of transcription initiation with nascent RNA distributions, in the main text we consider a model of transcriptional dynamics with an arbitrarily complex initiation mechanism followed by an elongation process. We make the following assumptions in our model (Fig. S1B): after initiation, each RNA polymerase molecule elongates along the gene by hopping from one to the next base at a constant probability per unit time. RNAP molecules do not pause and collide with other RNAP molecules, while moving along the gene. We also take the size of the RNAP footprint to be one base, and we do not restrict the number of RNAPs at each base along the gene. Below we give details of the calculations leading up to analytic formulas for the first two moments of the nascent RNA distribution predicted by this model.

To compute the first two moments of the nascent RNA distribution for an arbitrary promoter architecture, we consider a promoter that can exist in $N$ possible states. The rate of going from the $s$-th to the $q$-th promoter state is $k_{s,q}$, and the rate at which an RNAP molecule initiates transcription from the $s$-th promoter state is $k_{s,ini}$. After initiation every RNAP molecule elongates by hopping from one base to the next at a constant rate $k$ and the number of nascent RNAs at site $i$ is $m_i$. If the number of bases along the gene is $L$, then the number of nascent RNAs along the gene will be given by

$$M = \sum_{i=1}^{L} m_i. \qquad (1)$$

Here $m_i$'s are dependent random variables since at each base the number of nascent RNAs ($m_i$) depends on the number of nascent RNAs ($m_{i-1}$) present on the previous base. The state of the gene is described by ($L + 1$) stochastic variables: the number of nascent RNAs ($m_1,...,m_L$) at each base along the gene, and the state of the promoter, $s$. Thus, the probability distribution function describing the gene is given by $P(s,m_1,...,m_L)$. The time evolution for this probability can be calculated using the chemical master equation approach:

$$\frac{dP(s,m_1,...,m_L)}{dt} = k_{s,ini} P(s,m_1-1,...,m_L) + \sum_{i=1}^{L} k(m_i+1) P(s,m_1,.,m_i+1,m_{i+1}-1,.,m_L)$$

$$- (k_{s,ini} + \sum_{i=1}^{L} k\, m_i + \sum_q k_{s,q}) P(s,m_1,.,m_i,.,m_L) + \sum_q k_{q,s} P(s,m_1,...,m_L) \;. \qquad (2)$$

Similar approaches have been used previously in order to find the moments of mRNA and protein distributions (1–4).

By defining the probability vector $\vec{P}(m_1,...,m_L) = (P(1,m_1,...,m_L), P(2,m_1,...,m_L),...,P(s,m_1,...,m_L))$, we can rewrite Eq. 2 in compact matrix form as

$$\frac{d\vec{P}(m_1,...,m_L)}{dt} = (\hat{K} - \hat{R} - \hat{\Gamma}\sum_{i=1}^{L} m_i)\vec{P}(m_1,.,m_i,.,m_L) + \hat{R}\vec{P}(m_1-1,...,m_L)$$
$$+ \sum_{i=1}^{L} k(m_i+1)\hat{\Gamma}\vec{P}(m_1,.,m_i+1, m_{i+1}-1,.,m_L) \ . \tag{3}$$

Where we introduce the following matrices: $\hat{K}$, (which captures the transition between different promoter states) whose elements are $K_{qs} = k_{q,s}$, if $q \neq s$ and $K_{ss} = -\sum_q k_{q,s}$ ; $\hat{R}$ is a matrix that contains the rates of initiation from different promoter states. In the case of one-step initiation it is diagonal with the diagonal elements being the rates of initiation from different promoter states i.e. $R_{sq} = k_{s,ini}\delta_{s,q}$ . In the case of two-step initiation this matrix is off-diagonal owing to the fact that the promoter state changes after initiation. For instance if the promoter switches from the *s*-th to the *q*-th state after initiation, then the off-diagonal term that exists is, $R_{qs} = k_{s,ini}$. $\hat{\Gamma}$, which captures the hopping process of RNAP molecules from one base to the next, is diagonal and the corresponding matrix elements are $\Gamma_{sq} = k\delta_{s,q}$. In the steady state the left hand side of Eq. 3 is equal to zero.

**Mean**

From Eq. 1, the mean number $\langle M \rangle$ of nascent RNAs along the gene is given by

$$\langle M \rangle = \sum_{i=1}^{L} \langle m_i \rangle \ . \tag{4}$$

Hence to compute the mean of the steady-state probability distribution of nascent RNAs along the gene, we multiply both sides of Eq. 3 by $m_i$, and sum over all values of $m_i$ from 0 to $\infty$. We obtain two separate equations for $\langle m_1 \rangle$ and $\langle m_i \rangle$, where *i* goes from *2* to *L*.

For $m_1$ we get:

$$\sum_{m_1,...,m_L=0}^{\infty} m_1(\hat{K} - \hat{R} - \hat{\Gamma}\sum_{i=1}^{L} m_i)\vec{P}(m_1,.,m_i,.,m_L) + \sum_{m_1,...,m_L=0}^{\infty} m_1\hat{R}\vec{P}(m_1-1,...,m_L)$$
$$+ \sum_{m_1,...,m_L=0}^{\infty} \sum_{i=1}^{L} m_1 k(m_i+1)\hat{\Gamma}\vec{P}(m_1,.,m_i+1,m_{i+1}-1,.,m_L) = 0 \ . \tag{5}$$

Since none of the three matrices in Eq. 5 ($\hat{K}$, $\hat{R}$ and $\hat{\Gamma}$) are functions of $m_1$, they can be taken out of the sums. Simplifying the above equation further we find:

$$\hat{K} \sum_{m_1,\ldots,m_L=0}^{\infty} m_1 \vec{P}(m_1,\ldots,m_L) - \hat{R} \sum_{m_1,\ldots,m_L=0}^{\infty} m_1 \vec{P}(m_1,\ldots,m_L) + \hat{R} \sum_{m_1,\ldots,m_L=0}^{\infty} m_1 \vec{P}(m_1-1,\ldots,m_L)$$

$$-\hat{\Gamma} \sum_{m_1,\ldots,m_L=0}^{\infty} m_1^2 \vec{P}(m_1,\ldots,m_L) - \hat{\Gamma} \sum_{i=2}^{L} \sum_{m_1,\ldots,m_L=0}^{\infty} m_1 m_i \vec{P}(m_1,.,m_i,.,m_L)$$

$$+\hat{\Gamma} \sum_{m_1,\ldots,m_L=0}^{\infty} m_1 (m_1+1) \vec{P}(m_1+1, m_2-1,..,m_L) \quad (6)$$

$$+\hat{\Gamma} \sum_{i=2}^{L} \sum_{m_1,\ldots,m_L=0}^{\infty} m_1 (m_i+1) \vec{P}(m_1,.,m_i+1, m_{i+1}-1,.,m_L) = 0.$$

The equation above can be expressed in terms of the following partial moment vectors:

$$\vec{m}_1^{(0)} = \sum_{m_1,\ldots,m_L=0}^{\infty} m_1^0 \vec{P}(m_1,\ldots,m_L),$$

$$\vec{m}_1^{(1)} = \sum_{m_1,\ldots,m_L=0}^{\infty} m_1^1 \vec{P}(m_1,\ldots,m_L). \quad (7)$$

These partial moment vectors are useful quantitates, as they are related to the moments of the probability distribution of $m_1$. For example, the mean number of nascent RNAs at the first base is given by:

$$\langle m_1 \rangle = \sum_{s=1}^{N} \sum_{m_1,\ldots,m_L=0}^{\infty} m_1 \, p(s,m_1,\ldots,m_L) = \sum_{m_1,\ldots,m_L=0}^{\infty} m_1 \, p(1,m_1,\ldots,m_L) + \ldots + \sum_{m_1,\ldots,m_L=0}^{\infty} m_1 \, p(N,m_1,\ldots,m_L).$$

(8)

We simplify Eq. 6 and express it in terms of the partial moment vectors, defined above. In order to do that we re-arrange the third, sixth and seventh term using the fact that $m_i$'s are dummy variables. We also use the fact that the number of nascent RNA molecules at different bases can never fall below 0 (i.e. $\vec{p}(m_1,.,m_i=-1,.,m_L)=0$, where $i$ goes from 1 to $L$). Using Eq. 6 and Eq. 7 we obtain equations for the two partial moment vectors:

$$\hat{K} \vec{m}_1^{(0)} = 0,$$

$$(\hat{K} - \hat{\Gamma}) \vec{m}_1^{(1)} + \hat{R} \vec{m}_1^{(0)} = 0. \quad (9)$$

We can compute the mean of $m_1$, by multiplying Eq. 9 with $\hat{u} = (1,1,\ldots,1)$:

$$\langle m_1 \rangle = \frac{\vec{r} \, \vec{m}_1^{(0)}}{k}. \quad (10)$$

Here the vector $\vec{r}$ contains the ordered list of rates of transcription initiation from each promoter state. To compute the mean of the nascent RNA number at the other bases we multiply Eq. 3 by $m_i$, where $i$ goes from 2 to L.

$$\hat{K} \sum_{m_1,\ldots,m_L=0}^{\infty} m_i \vec{P}(m_1,.,m_i,.,m_L) - \hat{R} \sum_{m_1,\ldots,m_L=0}^{\infty} m_i \vec{P}(m_1,\ldots,m_L) + \hat{R} \sum_{m_1,\ldots,m_L=0}^{\infty} m_i \vec{P}(m_1-1,\ldots,m_L)$$

$$-\hat{\Gamma} \sum_{m_1,\ldots,m_L=0}^{\infty} m_i^2 \vec{P}(m_1,.,m_i,.,m_L) - \hat{\Gamma} \sum_{i \neq j} \sum_{m_1,\ldots,m_L=0}^{\infty} m_i m_j \vec{P}(m_1,.,m_i,.,m_j,.,m_L)$$

$$+\hat{\Gamma} \sum_{m_1,\ldots,m_L=0}^{\infty} m_i (m_i+1) \vec{P}(m_1,.,m_i+1,m_{i+1}-1.,m_L)$$

$$+\hat{\Gamma} \sum_{i \neq j} \sum_{m_1,\ldots,m_L=0}^{\infty} m_i (m_j+1) \vec{P}(m_1,.,m_i,.,m_j+1,m_{j+1}-1,.,m_L) = 0 \ .$$

(11)

As before, we define the following partials moment vectors for convenience:

$$\vec{m}_{i-1}^{(1)} = \sum_{m_1,\ldots,m_L=0}^{\infty} m_{i-1}^1 \vec{P}(m_1,.,m_{i-1},.,m_L) \ ,$$

$$\vec{m}_i^{(1)} = \sum_{m_1,\ldots,m_L=0}^{\infty} m_i^1 \vec{P}(m_1,.,m_i,.,m_L) \ .$$

(12)

Following the same procedure as $m_1$, we can find the mean of $m_i$ from Eq. 11, in terms of the partial moments (Eq. 12) of $m_i$ and $m_{i-1}$:

$$(\hat{K} - \hat{\Gamma}) \vec{m}_i^{(1)} + \hat{\Gamma} \vec{m}_{i-1}^{(1)} = 0 \ . \tag{13}$$

We compute the mean of $m_i$, by multiplying Eq. 13 with $\hat{u} = (1,1,\ldots,1)$. The transition matrix $\hat{K}$ has the property that the sum of the elements of any one of its columns is always 0. Using this property, we find that, $\vec{u}\,\hat{K} = 0$. Therefore, we find that the mean nascent RNA number at each base is the same as the mean of the previous one, i.e.,

$$\langle m_i \rangle = \langle m_{i-1} \rangle \ . \tag{14}$$

Using Eq. 12 and Eq. 13 we get:

$$\langle m_L \rangle = \langle m_{L-1} \rangle = \ldots = \langle m_1 \rangle \ . \tag{15}$$

In other words in the steady state, the mean number of nascent RNAs at every base along the gene is the same. Hence the mean nascent RNA number $\langle M \rangle$ along the gene is the sum of the mean of nascent RNA numbers at each base. Therefore from Eq. 1, Eq. 9 and Eq. 14, we get:

$$\langle M \rangle = L \frac{\vec{r}\,\vec{m}_1^{(0)}}{k} \ . \tag{16}$$

For instance, using this formula above one can find the mean number of nascent RNAs along a gene which initiates transcription in uncorrelated events at a constant rate *r*. For such a gene the mean number of nascent RNAs is given $\langle M \rangle = L \frac{r}{k}$ .

**Variance**

In order to compute the variance, we need to find all the elements of the covariance matrix of the dependent random variables, ($m_1, ... m_L$). Using Eq. 1, we find that the second moment is:

$$\langle M^2 \rangle = \sum_{i=1}^{L} \langle m_i^2 \rangle + 2 \sum_{j \neq i} \langle m_i m_j \rangle .$$

We expand it for convenience:

$$\langle M^2 \rangle = \langle m_1^2 \rangle + 2 \langle m_1 m_2 \rangle + 2 \sum_{i=3}^{L} \langle m_1 m_i \rangle + \sum_{i=2}^{L} \langle m_i^2 \rangle + 2 \sum_{i=2}^{L-1} \langle m_i m_{i+1} \rangle + 2 \sum_{j=i+2}^{L} \sum_{i=2}^{L-2} \langle m_i m_j \rangle . \qquad (17)$$

To find $\langle M^2 \rangle$, we need to compute each term on the right hand side using Eq. 3. As before we define the following partial moments:

$$\vec{m}_i^{(2)} = \sum_{m_1,...,m_L=0}^{\infty} m_i^2 \, \vec{P}(m_1,...,m_L) ,$$

$$\vec{m}_{i,j}^{(2)} = \sum_{m_1,...,m_L=0}^{\infty} m_i^1 \, m_j^1 \, \vec{P}(m_1,.,m_i,.,m_j,.,m_L) . \qquad (18)$$

Here *i* and *j* both range from 1 to *L*. We get a set of six equations for the partial moments defined above by multiplying Eq. 3 with $m_1^2$, $m_1 m_2$, $m_1 m_i$ (where *i* goes from 3 to *L*), $m_i^2$ (where *i* goes from 2 to *L*), $m_i m_{i+1}$ (where *i* goes from 2 to *L-1*) and $m_i m_j$ (where *i* goes from *2* to *L-2* and *j* runs from *i+2* to *L*)

$$(\hat{K} - 2\hat{\Gamma}) \vec{m}_1^{(2)} + 2\hat{R} \, \vec{m}_1^{(1)} + \hat{\Gamma} \, \vec{m}_1^{(1)} + \hat{R} \, \vec{m}_1^{(0)} = 0 . \qquad (19)$$

$$(\hat{K} - 2\hat{\Gamma}) \vec{m}_{1,2}^{(2)} - \hat{\Gamma} \, \vec{m}_1^{(1)} + \hat{\Gamma} \, \vec{m}_1^{(2)} + \hat{R} \, \vec{m}_2^{(1)} = 0 . \qquad (20)$$

$$(\hat{K} - 2\hat{\Gamma}) \vec{m}_{1,i}^{(2)} + \hat{\Gamma} \, \vec{m}_{1,i-1}^{(2)} + \hat{R} \, \vec{m}_3^{(1)} = 0 . \qquad (21)$$

$$(\hat{K} - 2\hat{\Gamma}) \vec{m}_i^{(2)} + 2\hat{\Gamma} \, \vec{m}_{i,i-1}^{(2)} + \hat{\Gamma} (\vec{m}_i^{(1)} + \vec{m}_{i-1}^{(1)}) = 0 . \qquad (22)$$

$$(\hat{K} - 2\hat{\Gamma}) m_{i,i+1}^{(2)} + \hat{\Gamma} \vec{m}_{i-1,i+1}^{(2)} + \hat{\Gamma} (\vec{m}_i^{(2)} - \vec{m}_i^{(1)}) = 0 \ . \tag{23}$$

$$(\hat{K} - 2\hat{\Gamma}) \vec{m}_{i,j}^{(2)} + \hat{\Gamma} \vec{m}_{i,j-1}^{(2)} + \hat{\Gamma} m_{i-1,j}^{(2)} = 0 \ . \tag{24}$$

From Eq. (19-24), we can compute the second order moments for the nascent RNA numbers at every base along the gene by multiplying the partial moment vectors with $\hat{u} = (1,1,...,1)$. In order to obtain the variance, we construct the covariance matrix, whose element in the *i, j* position is the covariance between the random variables characterizing the number of nascent RNAs at the *i*-th and *j*-th bases along the gene, defined as

$$\text{Covariance}(m_i m_j) = \langle m_i m_j \rangle - \langle m_i \rangle \langle m_j \rangle = \langle m_i m_j \rangle - \langle m_1 \rangle^2 \ . \tag{25}$$

Here we use the fact that $\langle m_i \rangle = \langle m_j \rangle = \langle m_1 \rangle$. The dimension of the covariance matrix is $L \times L$ and since the number of bases *L* along a gene is typically of the order of few thousands, the number of second order moments to be evaluated to construct the covariance matrix is very high. However, it is readily evaluated with Mathematica using Eq. (19-24) and Eq. 25. Therefore the variance of the number of nascent RNA molecules along a gene is given by

$$\text{Variance} = \langle M^2 \rangle - N^2 \langle m_1 \rangle^2 , \tag{26}$$

which is basically the sum of all the elements of the covariance matrix.

## Traffic jams and pauses during elongation do not affect the distribution of nascent RNAs in the slow initiation limit

Our model (Fig. S1B) makes the assumptions that RNAP molecules do not pause and do not collide with other RNAP molecules, while moving along the gene. We also take the size of the RNAP footprint to be one base, and we do not restrict the number of RNAPs at each base along the gene. These assumptions are equivalent to the assumption that the number of transcribing RNA polymerases is much less than one per base.

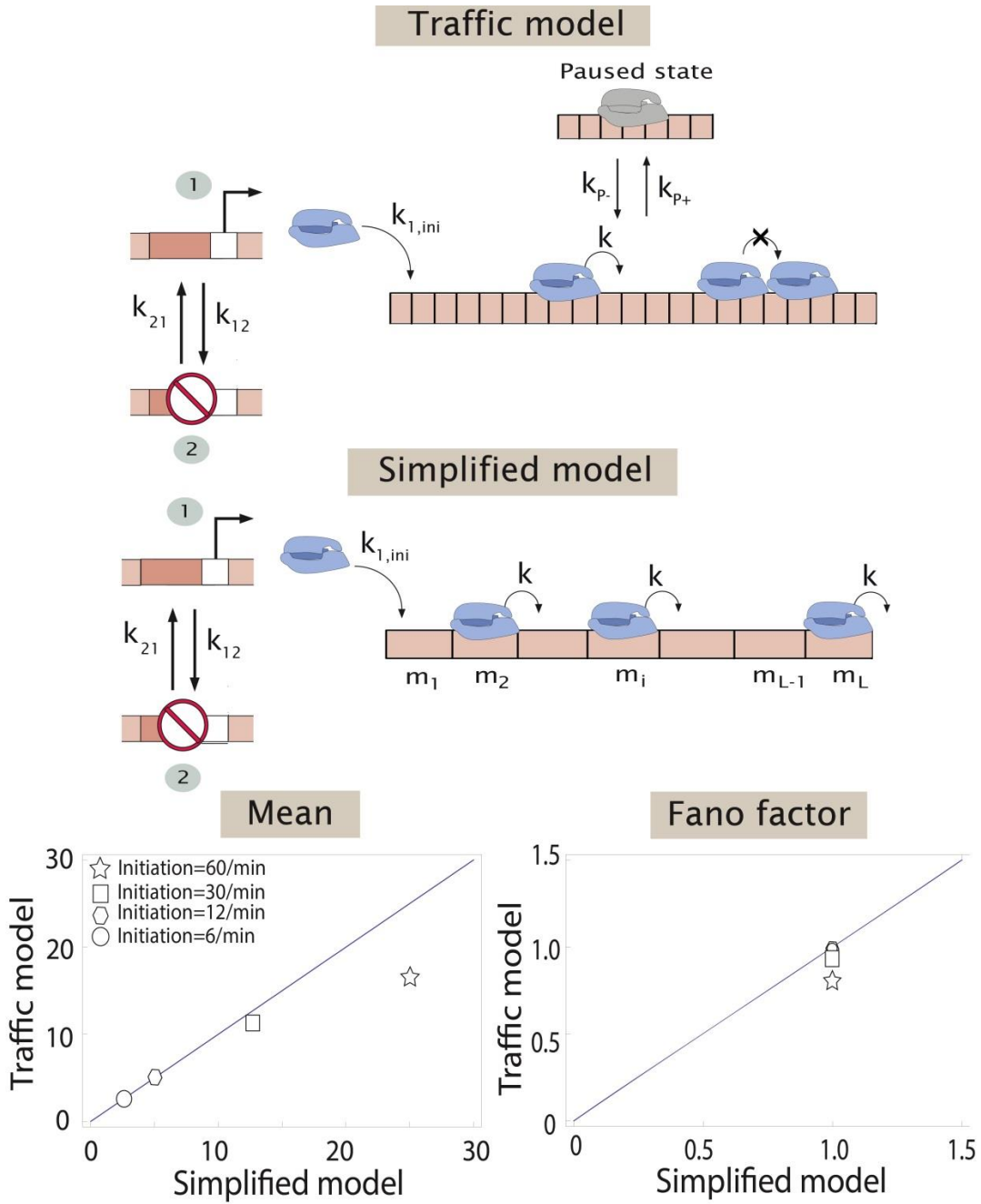

Fig. S1. (A) **Two state promoter with traffic**. The promoter switches between two states: state *1*, from which transcription initiation occurs with a constant probability per unit time $k_{1,ini}$, and state 2, from which transcription does not initiate. The promoter switches from state 1 to state 2 with probability per unit time $k_{12}$, and from 2 to 1 with probability per unit time $k_{21}$. After initiation each RNAP molecule hops from one base pair to the next along the gene at a rate $k$ per unit time. Each RNAP molecule has a finite DNA footprint of *30 bp* and it can pause at any site with a rate $k_{P+}$ and come out of the pause with a rate $k_{P-}$. An RNAP molecule cannot move forward if another one occupies the bases in front of it. The length of the gene is *L.* (B) **No-traffic model.** The elongation process is uniform with each RNAP molecule

occupying one base pair. (C) Using the Gillespie algorithm (7, 8), we simulate the traffic model where the promoter initiates at a constant rate, where we take four different values of initiation. Predictions of the no-traffic model for the mean and Fano factor agree well with the simulation results from the traffic model up to an initiation rate of 30 initiations/min. The elongation parameters used, are $k_{P-}$= 4/sec, $k_{P+}$=0.01/sec, $k$=80 bp/sec (5) as reported for ribosomal promoters in *E.Coli*. The length of the gene is *L*= 2000 bps and the footprint of one RNAP molecule is 30 bps.

The approximations made in our model are reasonable for all but the strongest promoters characterized by very fast initiation (5, 6). In order to test their validity, we compare the analytic predictions of our model with numerical simulations of a more detailed and realistic one (referred to as the traffic model (Fig. S1A)), which properly accounts for the footprint of a transcribing RNAP molecule on the DNA, ubiquitous pausing, and excluded volume interactions between adjacent polymerases along the gene. In particular we compare the mean and the Fano factor of nascent RNA distributions, as predicted by our model of transcription for the case when initiation occurs via a single rate limiting step, with those obtained from numerical simulation of the traffic model obtained using the Gillespie algorithm (7, 8).

A single time step of the simulation is performed in the following way: one of the set of all possible reactions is chosen according to its relative weight that depends on the rate, characterizing the step, and the state of the system is updated accordingly. The time elapsed since the last step is drawn from an exponential distribution, the rate parameter of which equals the sum of all the rates of the possible reactions at that time. This process is repeated iteratively for a long enough time such that the number of RNAP molecules along the gene (which is the same as the number of nascent RNAs) reaches steady state.

We consider four different transcription initiation rates, spanning the typically observed values in *E. coli* and yeast cells (9–12), and see how the mean and Fano factor of the nascent RNA distribution is affected by RNAP pausing and road blocking (Figs. S1A-B). We find that for initiation rates slower than 30 initiations/min, both the mean and the Fano factor extracted from the simulations are in good agreement (less than 10% difference) with the analytical results (Figs. S1C-D). All the initiation rates that have been reported in vivo are generally slower than 30 initiations/min, with important exceptions such as the ribosomal promoters (5). Hence the theory adopted in the main text, which does not include pausing and traffic jams by RNA polymerases, should be applicable to most genes in *E.Coli* and yeast. In simulations we used the following parameters to describe RNAP elongation: $k_{P-}$= 4/sec, $k_{P+}$=0.01/sec, $k$=80 bp/sec, as was reported for ribosomal promoters in *E.Coli* (5). We also use a gene of length *L*= 2000 bps and a polymerase whose DNA footprint is 30 bases.

## Methods

**Data analysis: Yeast genes**

We analyze the measured nascent RNA distributions for twelve different constitutively expressed yeast genes reported in reference (12). By applying our theoretical results to the published data, we find that for six of these twelve genes, the mean number of nascent RNAs scales linearly with the gene length; see Fig. S2, where the mean number of nascent RNAs is plotted against the gene length for all the twelve genes. If we assume that all of these genes have similar elongation rates (0.8 kb/min (10)), then the linear relationship between the mean nascent RNA number and gene length implies that the average initiation rates of the six (KAP104, TAF5, TAF6, TAF12, RPB2, RPB3) of these twelve genes are all roughly the same, and equal to 0.145±0.025/min. However the other six genes (RPB1, MDN1, PUP1, PRE7, PRE3, PRP8) initiate transcription at different rates.

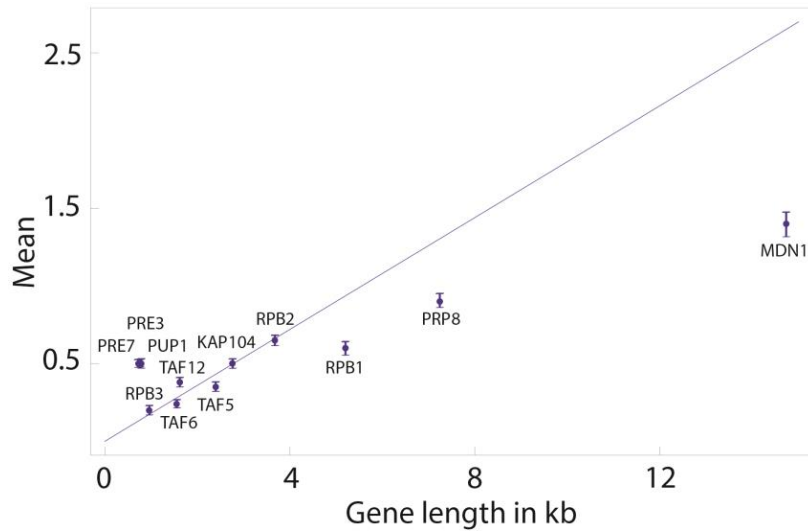

Fig. S2. **Mean number of nascent RNAs in yeast.** We use the nascent RNA distribution data for all the twelve genes: MDN1, PRP8, RPB1, PUP1, PRE3, PRE7, KAP104, TAF5, TAF6, TAF12, RPB2, RPB3, reported by Gandhi et al. (12). For six of the above twelve genes, KAP104, TAF5, TAF6, TAF12, RPB2, RPB3, the mean increases linearly with the gene length, as shown in the main text in Fig. 2A. The other six genes do not follow the same pattern indicating that they have different initiation rates (PUP1, PRE3, PRE7, have similar initiation rates though but different from the six genes analyzed in the main text.)(12).

In addition to the mean, we investigate the behavior of the Fano factor of the nascent mRNA distributions as well, and compare it with the prediction from our model of transcriptional regulation (Fig. 1A), as described in the main text. Since the Fano factors of the nascent mRNA distribution for all of the twelve genes in the data set is less than (or at most equal to) one, the simplest model consistent with the published data for these twelve yeast genes is one where the promoter is always active and transcription initiation is a two-step process (see Fig. 1A) parameterized by the kinetic rates $k_{OPEN}$ and $k_{ESC}$. The Fano factor of the distribution depends on the ratio of these two rates. For the six genes that initiate transcription at the same average rate, we plot in Fig. 2B the predicted region in the Fano factor– Gene Length phase space that is consistent with the observed average initiation rate of 0.145±0.025/min. This region (shaded in light blue in Fig.2B) is bounded on its upper side by the limit when one of the rates is much larger than the other one and on the lower side, by the Fano factor obtained when both rates are exactly identical. This line marks the minimum Fano factor attainable when the average initiation rate is 0.145±0.025/min. Remarkably, we find that the six genes: RPB2, RPB3, TAF5, TAF6, TAF12, KAP104, follow precisely the trend-line expected for when both, $k_{OPEN}$ and $k_{ESC}$ are the same ($k_{OPEN}=k_{ESC}=0.29±0.013$/min), as shown in Fig. 2B.

We also analyzed the Fano factor of the nascent mRNA distributions of the other six genes: RPB1, MDN1, PUP1, PRE7, PRE3, PRP8, which initiate transcription at a different rate (Fig. S2). Using the same procedure that we use to analyze the Fano factor for the other six genes, we find that in the Fano factor- Gene length phase space these six different genes are scattered, without a pattern like the other six genes, as shown in Fig. S3.

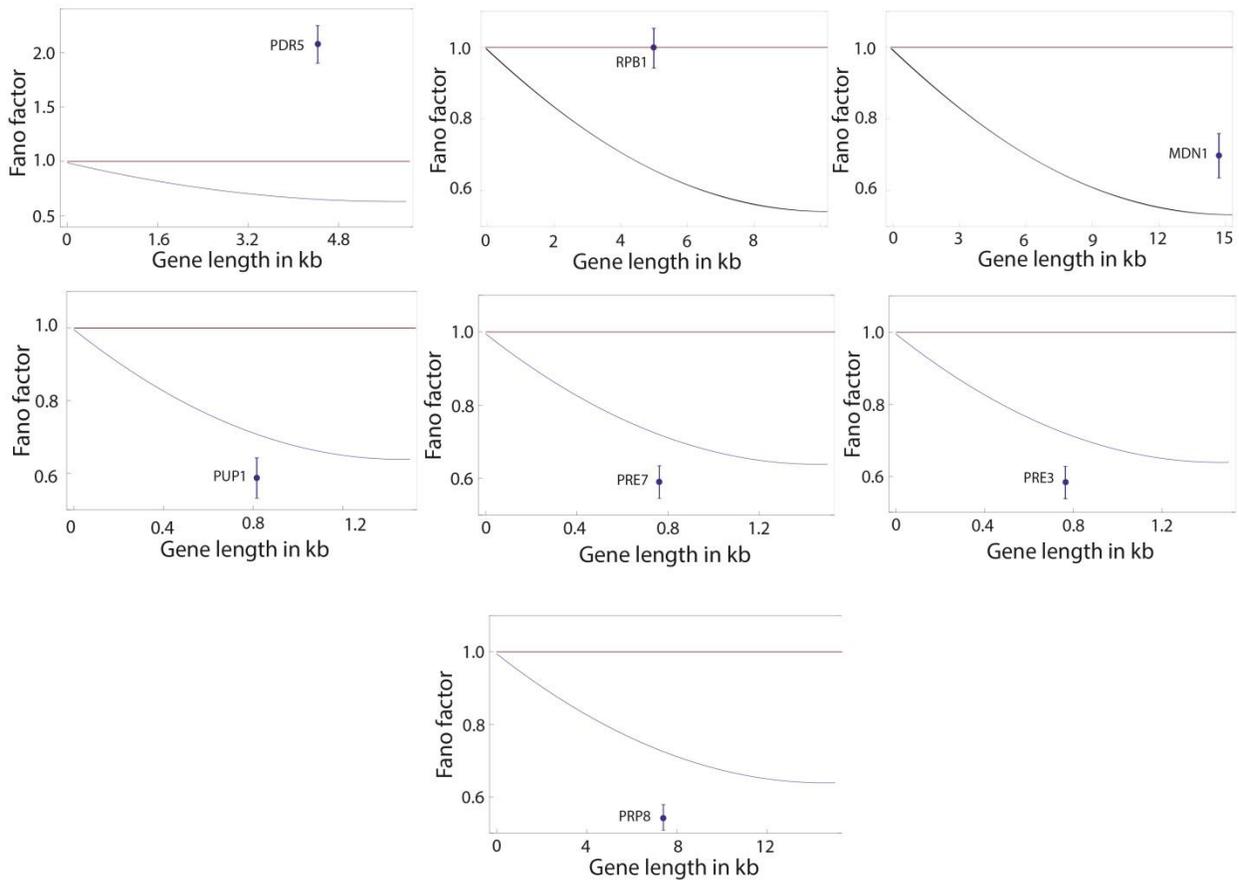

Fig. S3. **Fano factor for nascent RNA distributions in yeast.** Here we show 6 different constitutively expressed genes in yeast: RPB1, MDN1, PUP1, PRE7, PRE3, PRP8 (other than the six genes analyzed in the main text) (12) and another gene PDR5, known to be regulated (10). The data for Fano factors measured in experiments are shown in comparison with the predictions for the one-step (red line) and two-step (where both $k_{OPEN}$ and $k_{ESC}$ are the same; blue line) initiation models. The Fano factor for PDR5 gene is more than one which is consistent with an ON-OFF model of transcription initiation. RPB1 has a Fano factor equal to 1, suggesting a one-step initiation. All the other genes have Fano factors consistent with more than one rate limiting step leading up to initiation. The blue line indicates the minimum possible Fano factor for the specific gene assuming two-step initiation and the measured average initiation rate, which we compute from the mean number of nascent RNAs. For the genes whose Fano factor is below this line (PUP1, PRE7, PRE3, PRP8) the data suggest that their transcription is initiated via three or more (similar in duration) steps.

This suggests that not only these genes have different initiation rates; their mechanisms of initiation are also different. The Fano factor is one for one of these six genes and less than one for the other five. RPB1 has one step, MDN1 has two step and PUP1, PRE3, PRE7, PRP8 have more than two exponentially distributed in time steps leading up to initiation.